\documentclass{ws-p8-50x6-00}

\begin{document}

\title{Supernovae data: cosmological constant or ruling out the 
Cosmological Principle ?}

\author{M. N. C\'el\'erier}

\address{D\'epartement d'Astrophysique Relativiste et de Cosmologie, 
Observatoire de Parie-Meudon, 5 place Jules Janssen, 92195 Meudon 
C\'edex, 
FRANCE\\E-mail: Marie-Noelle.Celerier@obspm.fr}

\maketitle

\abstracts{Analysed in the framework of homogeneous FLRW models, 
the magnitude-redshift data from high redshift supernovae yield,
as a primary result, a strictly positive cosmological constant. 
Another reading of the currently published measurements does not 
exclude a possible ruling out of the Cosmological Principle 
and, thus, also, of the cosmological constant hypothesis. It is 
shown how shortly coming data can be used to settle this fondamental 
issue, pertaining to both cosmology and particle physics.}

\section{Introduction}

The discovery of high-redshift type Ia supernovae (SNIa) and their use 
as standard candles have resurected interest in the magnitude-redshift 
relation as a tool to measure the cosmological parameters of the 
universe. \\

Data recently collected by two survey teams (the Supernova Cosmology 
Project and the High-z Supernova Search Team), and analysed in the 
framework of homogeneous FLRW cosmological models, have yielded, as 
a primary result, a strictly positive cosmological constant, of order 
unity~\cite{ri,pe}. If these results were to be confirmed, it would be 
necessary to explain how $\Lambda$ is so small, yet non zero. Hence a 
revolutionary impact. \\

The purpose is here :\\

1. Assuming every source of potential bias or systematic uncertainties 
have been correctly taken into account in the data collecting, \\

2. Probe the large scale homogeneity of the region of the universe 
available with the SNIa measurements, thus testing the Cosmological 
Principle and cosmological constant hypotheses. \\

\section{Magnitude-redshift relation to probe large scale (in)homogeneity}

Consider any cosmological model for which the luminosity distance $D_L$ 
is a function of the redshift $z$ and of the parameters $cp$ of the 
model. Assume that $D_L$ is Taylor expandable near the observer, i.e. 
around $z=0$,

\begin{eqnarray}
&&D_L(z;cp)=\left(dD_L\over dz\right)_{z=0}z \, + \, {1\over 2}
\left(d^2D_L\over dz^2\right)_{z=0}z^2  \nonumber \\
&&+ \, {1\over 6}\left(d^3D_L \over dz^3\right)_{z=0}z^3 \, + \, 
{1\over 24}\left(d^4D_L\over dz^4\right)_{z=0}z^4 + {\cal O}(z^5) . 
\label{eq:1}
\end{eqnarray}

The apparent bolometric magnitude $m$ of a standard candle of absolute 
bolometric magnitude $M$ is also a function of $z$ and $cp$. In 
megaparsecs,

\begin{equation}
m=M+5 \log D_L(z;cp)+25.
\label{eq:2}
\end{equation}

Luminosity-distance measurements of such sources at increasing redshifts 
$z<1$ thus yield values for the coefficients at increasing order in the 
above expansion. For cosmological models with high, or infinite, 
number of free parameters, the observations only produce constraints 
upon the parameter values near the observer. For cosmological models 
with few constant parameters, giving independent contributions to each 
coefficient in the expansion, the observed magnitude-redshift relation 
provides a way: \\

1. To test the validity of the model. \\

2. If valid, to evaluate its parameters. \\

For Friedmann models precisely, the expansion coefficients $D_L^{(i)}$ 
are independent functions of the three parameters $H_0$, $\Omega_M$ and 
$\Omega_\Lambda$, and can be derived from the well-known expression 
of $D_L$~\cite{ce}. Therefore, accurate luminosity-distance measurements 
of three samples of same order redshift SNIa - one at $z\sim 0.1$, one 
at $z\sim 0.5$ and one at $z\sim 0.7$, for instance - would yield 
values for $D_L^{(1)}$, $D_L^{(2)}$ and $D_L^{(3)}$ and 
thus select a triplet of numbers for the model parameters $H_0$, 
$\Omega_M$ and $\Omega_\Lambda$. \\

Would the value of $\Omega_M$, in this 
triplet, be negative, and thus physically inconsistent - which cannot be 
excluded from the current data - the Friedmann cosmology would have to be 
ruled out at this stage. Would this value be positive, the triplet could 
be used to provide a prediction for the value of the forth order 
coefficient $D_L^{(4)}$. Now, if further observations at redshifts 
approaching unity could be made - $z\sim 0.8-0.9$ would suffice for a 
measurement accuracy of order 5-10\% - $D_L^{(4)}$ could be determined 
and compared to its predicted value, thus providing a test of the FLRW 
model. \\

If the ongoing surveys were to discover more distant sources, at 
redshifts higher than unity, the Taylor expansion would no longer be 
valid. One would have to consider numerical methods to select the 
theoretical model best fitting the data and complete the test of the 
homogeneity assumption~\cite{ce}.

\section{Example of alternative inhomogeneous model of universe}

The ruling out of the FLRW paradigm and of the related Cosmological 
Principle is not a purely academical possibility. Physically robust 
inhomogeneous models exist, which can verify any observed 
magnitude-redshift relation. Furthermore, a non-zero cosmological 
constant is not mandatory, as $\Lambda=0$ inhomogeneous models can mimic 
$\Lambda\neq 0$ Friedmann ones. \\

Lema\^itre-Tolman-Bondi (LTB) models~\cite{le,to,bo} are spatially 
spherically symmetrical solutions of Einstein's equations with dust 
as source of gravitationnal energy. They can thus be retained to 
roughly represent a quasi-isotropic universe in the matter dominated 
area. \\

Einstein's equations with $\Lambda=0$ imply that the metric 
coefficients, in proper time and comoving coordinates, are functions 
of the time-like $t$ and radial $r$ coordinates, and of two 
independent functions of $r$, which play the role of model 
parameters. The radial luminosity distance $D_L$ can be 
expressed as a function of $t$, $r$, the redshift $z$ and the two 
above cited independent functions of $r$. In the approximation of 
a centered observer, the $D_L$ expansion coefficients follow, as 
independent functions of the derivatives of the model parameters, 
evaluated at the observer ($z=0$). These parameters, which are implicit 
functions of $z$ through the null geodesic equations, are present 
in each coefficient $D_L^{(i)}$ with derivatives of order $i$~\cite{ce}. 
LTB models are thus {\it completly degenerate with respect to any 
magnitude-redshift relation}. \\

One can therefore fit any observed relation with a class of 
$\Lambda=0$ LTB models fulfilling the constraints on its parameters 
proceeding from the data. In fact, a non-zero $\Lambda$ can also be 
retained in these models. This only adds a new free parameter in 
the equations, increasing the degeneracy of the models with respect 
to magnitude-redshift relations. It is in particular the case for 
the class of relations selected by the current SNIa measurements, which 
can be interpreted as implying either a non-zero cosmological constant 
in a FLRW universe, or large scale inhomogeneity with no constraint on 
$\Lambda$.

\section{Conclusions}

Provided SNIa would be confirmed as good standard candles, data from 
this kind of sources at redshifts approaching unity could, in a near 
future, be used to test the homogeneity assumption on our past light 
cone. \\

Using, as an example, the LTB solutions, it has here been shown that: \\

- would this assumption be discarded by the shape of the measured 
magnitude-redshift relation, inhomogeneous solutions could provide 
good alternative models, as they are completly degenerate with respect 
to any of these relations, even with a vanishing cosmological 
constant. \\

- would a FLRW type distance-redshift relation be observed, it would 
not be enough to strongly support the Cosmological Principle. Even 
if this would imply a fine tuning for its parameters, the possibility 
for an inhomogeneous model to mimic such a relation could not be 
excluded. \\
 
Therefore, at the current stage reached by the observations, a 
non-zero $\Lambda$ is not mandatory, as, for example, a class of 
$\Lambda=0$ LTB models can mimic a $\Lambda\neq 0$ FLRW M-R relation. \\

In any case, to consolidate the robustness of future 
magnitude-redshift tests, it would be worth confronting their results 
with the full range of available cosmological data, {\it analysed in 
a model independent way}.


\begin{thebibliography}{99}
\bibitem{ri}A.G. Riess {\it et al}, \Journal{AJ}{116}{1009}{1998}.

\bibitem{pe}S. Perlmutter {\it et al}, \Journal{ApJ}{517}{565}{1999}.

\bibitem{ce}M.N. C\'el\'erier, A\&A (2000) in press, 
astro-ph/9907206.

\bibitem{le}G. Lema\^itre, \Journal{Ann. Soc. Sci. Bruxelles}
{A53}{51}{1933}.

\bibitem{to}R.C. Tolman, \Journal{Proc. Nat. Acad. Sci.}{20}{169}{1934}.

\bibitem{bo}H. Bondi, \Journal{MNRAS}{107}{410}{1947}.


\end{thebibliography}
\end{document}